\documentclass[twocolumn,trackchanges]{aastex62}
\usepackage{amsmath}
\usepackage{graphicx}
\usepackage{amsfonts}
\usepackage{natbib}
\usepackage{color}
\usepackage{epstopdf}
\usepackage{graphicx}
\usepackage{multirow}
\usepackage{longtable}
\usepackage{rotating}

\shorttitle{Gamma-ray periodicity in 47 Tuc}
\shortauthors{Zhang et al.}
\begin{document}

\title{A gamma-ray periodic modulation in Globular Cluster 47 Tucanae}

\author{Peng-Fei Zhang}
\affil{Department of Astronomy, School of Physics and Astronomy, Key Laboratory of Astroparticle Physics of Yunnan Province, Yunnan University, Kunming 650091, People’s Republic of China; zhangpengfei@ynu.edu.cn; fangjun@ynu.edu.cn}
\affil{Key Laboratory of Dark Matter and Space Astronomy, Purple Mountain Observatory, Chinese Academy of Sciences, Nanjing 210034, China; yzfan@pmo.ac.cn}

\author{Jia-Neng Zhou}
\affil{Shanghai Astronomical Observatory, Chinese Academy of Sciences, 80 Nandan Road, Shanghai 200030, People’s Republic of China; zjn@shao.ac.cn}

\author{Da-Hai Yan}
\affil{Key Laboratory for the Structure and Evolution of Celestial Objects, Yunnan Observatory, Chinese Academy of Sciences, Kunming 650011, People’s Republic of China}
\affil{Center for Astronomical Mega-Science, Chinese Academy of Sciences, 20A Datun Road, Chaoyang District, Beijing 100012, China}
\affil{Key Laboratory for the Structure and Evolution of Celestial Objects, Chinese Academy of Sciences, Kunming 650011, China}

\author{Jing-Zhi Yan}
\affil{Key Laboratory of Dark Matter and Space Astronomy, Purple Mountain Observatory, Chinese Academy of Sciences, Nanjing 210034, China; yzfan@pmo.ac.cn}
\affil{School of Astronomy and Space Science, University of Science and Technology of China, Hefei, Anhui 230026, China}

\author{Yi-Zhong Fan}
\affil{Key Laboratory of Dark Matter and Space Astronomy, Purple Mountain Observatory, Chinese Academy of Sciences, Nanjing 210034, China; yzfan@pmo.ac.cn}
\affil{School of Astronomy and Space Science, University of Science and Technology of China, Hefei, Anhui 230026, China}

\author{Jun Fang}
\affil{Department of Astronomy, School of Physics and Astronomy, Key Laboratory of Astroparticle Physics of Yunnan Province, Yunnan University, Kunming 650091, People’s Republic of China; zhangpengfei@ynu.edu.cn; fangjun@ynu.edu.cn}

\author{Li Zhang}
\affil{Department of Astronomy, School of Physics and Astronomy, Key Laboratory of Astroparticle Physics of Yunnan Province, Yunnan University, Kunming 650091, People’s Republic of China; zhangpengfei@ynu.edu.cn; fangjun@ynu.edu.cn}



\begin{abstract}

The Globular Cluster 47 Tucanae was firstly detected in gamma-rays by the Large Area Telescope (LAT)
onboard the \emph{Fermi} Gamma-ray Space Telescope,
and the gamma-ray emission has been widely attributed to the millisecond pulsars.
In this work, we analyze the Fermi-LAT pass 8 data ranging from 2008 August to 2017 May and report
the detection of a modulation with a period of $18.416\pm0.008$ hours at a significance level of $\sim 4.8\sigma$.
This is the first time to detect a significant modulation with a period much longer than that of millisecond pulsars in gamma-rays from Globular Clusters.
%\note{\sout{We also find significant variabilities in the monthly gamma-ray light curve of 47 Tucanae.}}
The periodic modulation signal appears in the {\it Swift}-BAT data as well. The phase-folded Chandra X-ray light curve of a point source
%(right ascension = $00^{\rm h}24^{\rm m}06^{\rm s}.389$, declination = $-72^{\circ}04'43''.006$)
may have provided an additional clue.
%\del{These facts indicate the detection of the first young pulsar binary in gamma-rays in Globular Clusters.}

\end{abstract}



\section{Introduction}
\label{sec:intro}

With ages greater than $10^{10}$ yr, Globular Clusters (GCs) are the oldest {spherical, self-gravitating aggregations of stars} orbiting the bulge of a host galaxy.
More than 150 GCs are detected in the Milky Way with radio and/or optical detectors \citep{Harris1996}.
47 Tucanae (henceforth 47 Tuc), located at a distance of $\sim4.0$ kpc from the Earth, was firstly detected
in gamma-rays by \emph{Fermi}-LAT \citep{Abdo2009a,Atwood2009}. The spectral energy distribution (SED) is similar to that of millisecond pulsars
(MSPs) \citep{Abdo2009a,Grindlay2001,Freire2011}.
Interestingly, 25 MSPs have been identified in 47 Tuc and the total number is estimated $>30$ \citep{Grindlay2001,Heinke2005b,Abdo2010}.
So far 16 gamma-ray GCs and
several candidates have been reported, and all the gamma-ray spectra resemble those of MSPs \citep{Acero2015,Kong2010,Abdo2010,Tam2011a,Zhou2015,Zhang2016}.
Together with the detection of {millisecond gamma-ray pulsations from a young MSP J1823$-$3021A in NGC 6624 \citep{Freire2011} and PSR B1821$-$24 in M28 \citep{Wu2013}},
these facts strongly favor the hypothesis that the GeV emission of 47 Tuc
or even all GCs mainly comes from a large population of MSPs.

GCs likely host other kinds of gamma-ray emitters as well.
%GCs are composed of a large number of stars.
Owing to the high stellar density and hence the large stellar encounter rate,
GCs are expected to host low mass X-ray binaries (LMXB, the progenitors of MSPs) \citep{Clark1975,Alpar1982,Cheng2003,Liu2007} and binary MSPs.
{Several binary MSPs, including Black Widow and Redback binary systems, are identified as gamma-ray emitters \citep{Guillemot2012,Wu2012,XingY2015}.
So far there are 15 known MSP binaries, including 5 quiescent-LMXBs and 1 black hole-LMXBs, already detected in 47 Tuc by radio,
optical and X-ray telescopes \citep{Heinke2005b,Miller-Jones2015}}.
A new gamma-ray emission component {with periodic modulation of hours} may thus present,
which motivates us to re-analyze the gamma-ray emission from 47 Tuc.

\section{Fermi-LAT data analysis}
\label{data}

We construct the LAT light curve using a modified version of aperture photometry with 500 s time-bin for 47 Tuc. We extract events within a circular region with a radius of $\sim3^\circ$ centered on the location of 47 Tuc in the energy range from 100 MeV to 500 GeV.
Using Tool \emph{gtmktime}, we select the good time intervals (GTIs) to exclude the periods when 47 Tuc was close to the Earth's limb or within 5 degrees of the Sun and the Moon.
Because a time-bin of 500 s is shorter than the sky survey rocking period of \emph{Fermi}-LAT, the large exposure changes from time-bin to time-bin for 47 Tuc.
So we use Tool \emph{gtexposure} to determine exposure for each time-bin. Then photon arrival times are barycenter corrected using Tool \emph{gtbary}.
We also use Tool \emph{gtsrcprob} to assign probabilities for each event basing on the new generated model file.
The light curve is calculated by summing the probabilities of events {(rather than the number of photons to remove the possible contamination of nearby sources)}
and weighting the relative exposures of each time-bin. Similar approaches are widely adopted in the literature \citep{Corbet2007,Abdo2009b,Ackermann2012}.
{The details of our data reduction are introduced in the Appendix.}

%%%%%%%
\begin{figure*}
\centering
\includegraphics[scale=0.5]{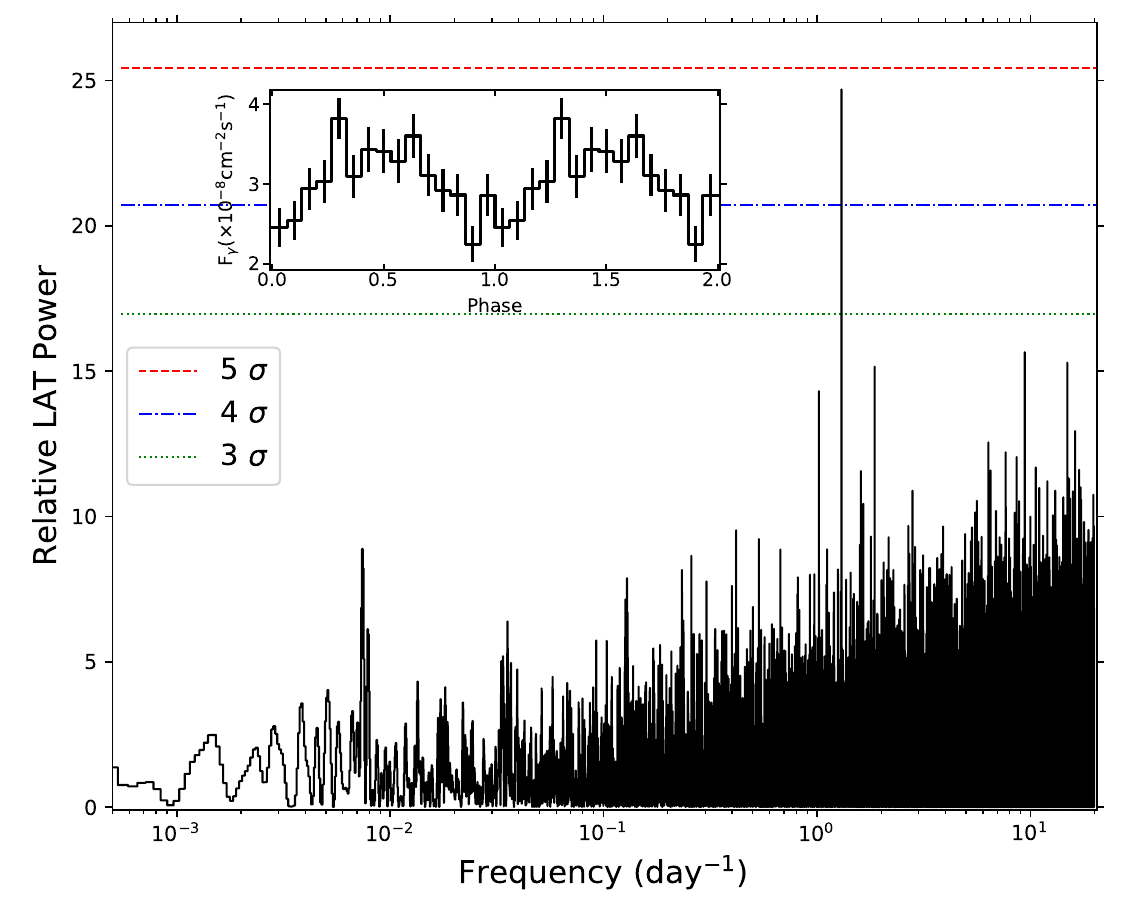}
\caption{Power spectrum calculated from gamma-ray light curve of 47 Tuc.
         The light curve is obtained with the aperture photometry method for the data above 100 MeV with the time-bin of 500 s.
         The green dotted, blue dashed-dotted, and red dashed horizontal lines represent
         $3\sigma$, $4\sigma$, and $5\sigma$ confidence levels, respectively. {The inset plot shows the phase-resolved light curves folded with its period.}}
\label{Fig1}
\end{figure*}
%**********************************Figure 2***************************************
\begin{figure*}
\centering
\includegraphics[scale=0.5]{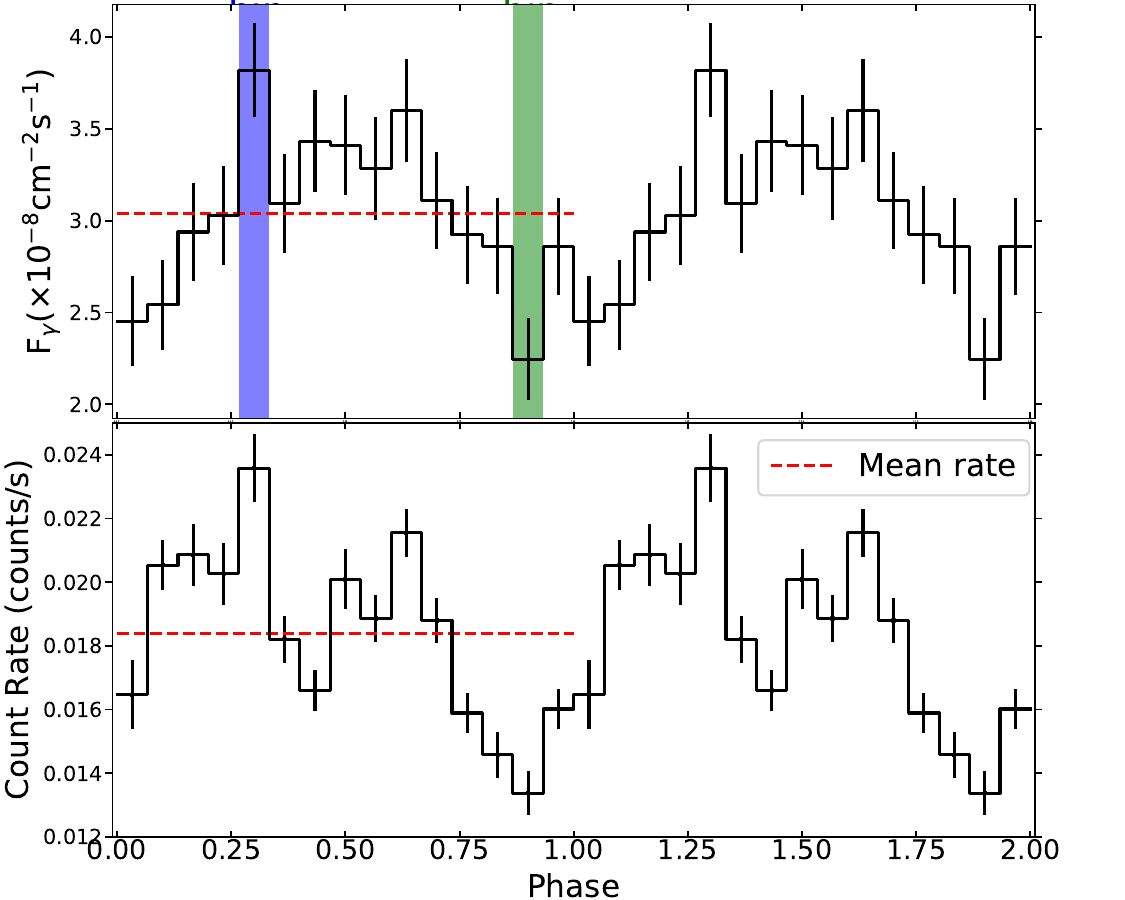}
\caption{Phase-resolved light curves (two cycles are shown for clarity).
              Upper panel: the gamma-ray flux intensity pulse shape folded with a period cycle of 18.416 hours. The red dashed line marks the mean flux.
              The phase-bins in high and low states are labeled with blue and green vertical bars, respectively.
              The highest and lowest TS values of the bins are $\sim$549 and $\sim$296, respectively.
              Lower panel: phase-resolved count rate with data from Chandra X-ray observatory. 
              %The strong modulation and similar shape displayed in gamma-ray and X-ray are caused by the orbital modulation of a new source.
              }
\label{Fig2}
\end{figure*}
%%%%%%%

We calculate the power spectrum for the entire light curve after subtracting the mean flux to examine the possible periodic modulation
{with method of Lomb-Scargle periodogram \citep{Lomb1976,Scarle1982}}.
%The power spectrum covers a period range from 0.05 day to the length of light curve (i.e., 3200 days).
The power spectrum is shown in Fig.~\ref{Fig1} and displays a distinct peak at $1.3032\pm0.0003~{\rm day}^{-1}$ ($\sim18.42$ hours).
The probability \citep[$p_{\rm prob}$; ][]{Lomb1976,Scarle1982} of obtaining power from the chance fluctuation (the noise) equal to or larger than power spectral peak is $<1.8\times10^{-11}$.
Taking into account the number of statistically independent frequency trials ($N$), the false-alarm probability (FAP) is $<1.15\times10^{-6}$.
And the FAP \citep{Scarle1982} is calculated as ${\rm FAP} =1-(1 - p_{\rm prob})^{N}$.
The significance of power spectral peak is estimated to be at $\sim4.8\sigma$ (99.999885\%) confidence level.
In Fig.~\ref{Fig1}, we also show $3\sigma$, $4\sigma$ and $5\sigma$ confidence levels with green dotted, blue dashed-dotted and red dashed lines, respectively.

{ As shown in the panel A of {Appendix} Fig.~\ref{fig:phase-resolved SED}, there is no significant emission above 50 GeV for 47 Tuc.
To increase the signal-to-noise of target, we re-analyzed the Fermi-LAT events with data up to 50 GeV, and the significance increases up to 4.9$\sigma$.
More intriguingly, if we limit the data up to $\sim$5 GeV, the significance increases to 5.2$\sigma$ and the power spectrum is shown in {Appendix} Fig.~\ref{Fig1-2}.}
To check whether our signal detected in 47 Tuc is artificial or not,
we also extract light curves with the same method for bright gamma-ray sources around 47 Tuc in 3FGL \citep{Acero2015}, then calculate their power spectra.
No similar signal has been identified.
% significant peak at the orbital period of 47 Tuc displays in their power spectra.
We further analyze the data of LS 5039, LS I +61$^{\circ}$303, Cygnus X-3, 1FGL J1018.6$-$5856,
and CXOU J053600.0$-$673507 and do confirm the signals (including their periods and the significance) reported in the literature
\citep{Abdo2009d,Abdo2009e,Abdo2009b,Ackermann2012,Corbet2016}.
%Moreover, we also test the method for estimating significance reported here, which is also well in agreement with confidence curves reported in the literature.
We then conclude that our signal is robust and create phase-resolved fluxes {employing likelihood analysis} within each phase-bin for gamma-rays (see the upper panel of Fig.2).

\section{SUMMARY AND DISCUSSION}
\label{4}

Gamma-ray orbit modulation has been detected in some high-mass X-ray binaries \citep[HMXBs; ][]{Abdo2009d,Tam2011b,Abdo2011,Ackermann2012,Corbet2016}.
Our localization of the emission in the high state yields an error circle of $\sim0.034^{\circ}$ that covers 47 Tuc. 
The superposition of a HMXB
in such a small field of view is thus unlikely, and we do not discuss such a scenario further.

An alternative and attractive possibility is that one of the {MSP binaries (either Redback or Black Widow system)} in 47 Tuc radiates significantly in gamma-rays.
The orbital periods of the 15 {MSP binaries} in 47 Tuc \citep{Heinke2005b,Miller-Jones2015}, inferred from radio and optical observations, span several hours to several days.
%Our gamma-ray modulation period $\sim18.4$ hours is indeed within such a range (see {Appendix} Fig.~\ref{fig:distribution}).
Our gamma-ray modulation period $\sim18.4$ hours is indeed within {such a range}.
However, none of the currently known {MSP binaries} in 47 Tuc has an orbital period of $\sim18.4$ hours, 
possibly indicating the presence of more {MSP binaries} in this source.

To further check this possibility and also identify the possible counterpart for the modulation, we have analyzed the {\it Swift}-BAT data, 
and foud a signal with almost the same orbital period as the gamma-ray modulation at a local significance level of $\sim 4.5\sigma$ (see the {Appendix} and the {Appendix} Fig.~\ref{Fig3}).
We have analyzed the Chanrda X-ray data from two Chandra ACIS detectors, and found an interesting source at coordinates of right ascension (R. A.) = $00^{\rm h}24^{\rm m}06^{\rm s}.389$
and declination (decl.) = $-72^{\circ}04'43''.006$ {(i.e., CXOGlb J002406.3-720443; see the {Appendix} and {Appendix} Fig.~\ref{Fig4})}.
%{\sout{The phase-folded light curve has a shape similar to that in gamma-rays and the peaks in these two bands are the same (see Fig.~\ref{Fig2}).}}
%Since the Chandra X-ray observations are not dense enough and the longest individual visit is less than 20 hours,
%the reliable identification of a $\sim 18.4$ hours periodic modulation single in the power spectrum analysis is rather challenging.
{The Chandra X-ray observations are not dense enough and the longest individual observation is less than 20 hours.
Consequently, the power spectrum at the frequencies of interest is not well defined with such short observations. 
 We therefore do not display its power spectrum analysis.}
%Nevertheless, we have tried to analyze the observations ranging from 2002 September 29 and 2002 October 11,
%in total four observation periods lasting $\sim73.5$ hours and found a peak at $21.6\pm8.3$ hours (see the {Appendix} Fig.~\ref{Fig5a}).
The X-ray folded light-curve {after removing the influencing of two flares} is shown in the lower panel of Fig. 2. Interestingly, {the phase-resolved gamma-ray and X-ray light curves have a similar shape.}
More importantly, the folded gamma-ray and X-ray light curves both reach their maximum in the 5th phase-bin.
We thus take CXOGlb J002406.3-720443 as the candidate of the source of our periodic gamma-ray signal.
Motivated by these {facts}, we conclude that there is a new gamma-ray emitter, possibly a {MSP binary}, in 47 Tuc.

Note that PSR 1957+20 \citep{Wu2012} and PSR J1311-3430 \citep{XingY2015}, two Black Windows, show orbital modulations in gamma-rays.
%Such orbital modulated gamma-ray emission might be due to the inverse Compton scattering between relativistic pulsar wind and soft photons from the companion star
{The orbital phase offset between X-ray and gamma-ray modulations suggests that their gamma-rays are produced by inverse Compton scattering \citep{Cheng2006,Tam2011b,Ng2018,Clark2020}, 
rather than the synchrotron of electrons which is considered as the origin of the X-ray radiations\citep{Bogdanov2005,An2018,An2020}.
While in 47 Tuc, the gamma-ray orbital modulation has the same orbital phase as the X-ray modulation, suggesting that 
this gamma-ray component from 47 Tuc is produced by synchrotron emission from particles accelerated along the shock front, same as the origin of the X-ray emissions.

The phase-resolved spectral properties and fluxes for maximum (phase-bin 5) and minimum (phase-bin 14) are listed in {Appendix} Table 1.
We find that the maximum spectrum ($\Gamma_{\rm H}~=~1.82\pm0.13$) is softer than minimum one  ($\Gamma_{\rm L}~=~1.26\pm0.25$). The difference of $\Gamma$ is $0.56\pm0.28$.
This suggests the orbitally varying component expected in the intra-binary shock emission models, and the similar scenario is also discussed in \citet{An2018}.}
%Furthermore, our results can not rule out the occultation of the gamma-rays from pulsar in binary system by the companion star.}
Dedicated X-ray and optical observations are encouraged/needed to finally reveal the nature of the gamma-ray periodic modulation in 47 Tuc.
%%%%%%%%%%%%%%%%%%%%5%%
%Acknowledgements %
%%%%%%
\section*{Acknowledgements}
{We thank the anonymous referee for the useful and constructive comments.
This work is supported in part by the National Key R\&D Program of China under grant No. 2018YFA0404204, the National Natural Science Foundation of China (
%No. 11525313, No. 11433004, No. 11433009,
No. 11603059,  No. 11661161010,
%No.11803081,
No. U1738124, and No. E085051002), and the joint foundation of Department of Science and Technology of Yunnan Province and Yunnan University $[$2018FY001 (-003)$]$, and the Candidate Talents Training Fund of Yunnan Province (2017HB003).
The work of D. H. Yan is also supported by the CAS Youth Innovation Promotion Association.}

\appendix
%\section{Autocorrelation}


{\bf Gamma-ray data reduction.} {In the analysis we employ the Science Tools available from the \emph{Fermi} Science Support Center (FSSC)
with software version v10r0p5 package.
The Fermi-LAT \citep{Atwood2009} data used here were observed between 2008 August 8 and 2017 May 20 (modified Julian date, MJD: 54,682.66-57,894.00).
We only select photons between 100 MeV and 500 GeV to reduce diffuse emission and improve point spread function.
We exclude the events with zenith angles $>90^{\circ}$ to minimize the contamination from the gamma-ray-bright Earth limb.
The high-quality data (Pass 8 SOURCE class photon-like events with options of \emph{evclass=128} and \emph{evtype=3})
are used to obtain the GTIs by running
\emph{Fermi} Science Tools \emph{gtmktime} with filter expression of ``(DATA\_QUAL$>$0)\&\&(LAT\_CONFIG==1)''.
The instrumental response function ``P8R2\_SOURCE\_V6'', recommended by \emph{Fermi}-LAT Collaboration, is adopted.

Our region of interest (ROI) is an area of $20^{\circ}\times20^{\circ}$ centered at position of 47 Tuc
%with right ascension (R. A.) = $00^{\rm h}24^{\rm m}05^{\rm s}.36$ and declination (decl.) = $-72^{\circ}04'53''$
(J2000; $l=305.895,~b=-44.889$).
We bin the data into $200\times200$ grid points with spatial pixel size of $0.1^{\circ}\times0.1^{\circ}$ in spatial dimensionality
and into 30 logarithmically equal bins in energy dimensionality.
Two latest background components (i.e., gll\_iem\_v06.fits and iso\_P8R2\_SOURCE\_V6\_v06.txt) are adopted to model gamma-rays
from Galactic diffuse and extragalactic isotropic emissions.
The background sources in the ROI are adopted from the Fermi LAT 4-Year Point Source Catalog {(3FGL) \citep{Acero2015}}.
A binned maximum likelihood algorithm packaged in Tool \emph{gtlike} is performed to fit the events with the model file.
In the modeling, we free the normalizations and spectral parameters of sources within the radius of $5^{\circ}$,
fix the spectral parameters to that reported in 3FGL {\citep{Acero2015}} for the sources at the radii between $5^{\circ}$ and $10^{\circ}$,
and freeze all the parameters for other sources. The exceptions are the highly variable sources with a variability index exceeding 72.44, for which
the normalizations are set free. So are the two background components.
For 47 Tuc, the spectral shape has a function of a super exponentially cutoff power law
%(PLSuperExpCutoff,
($dN/dE=N_0E^{-\Gamma}\exp(-E/E_{\rm cut})^b$; where the parameters of $N_0$, $E_{\rm cut}$, $\Gamma$, and $b$ are the normalization factor,
the cutoff energy, the photon index, and the sharpness of the cutoff, respectively).
Running the Tool \emph{gtlike}, we derive the best-fit results, including flux intensities, photon spectral indices, and likelihood ratio test statistic (TS) values.
The TS is defined as $\rm TS=-2(\mathcal{L}_{0}-\mathcal{L}_{1})$, where $\mathcal{L}_{0}$ and $\mathcal{L}_{1}$ represent logarithmic maximum likelihood values
under the null and alternative model (i.e., without and with the target).
The best-fit results are saved as a new model file, which is the base of the following data analysis.
Unless otherwise stated, the uncertainties given in this work are the $1\sigma$ statistical errors.

{\bf Phase-resolved gamma-ray emission.} To investigate the variability on the period cycle of 47 Tuc,
we employ the phase-resolved likelihood analysis method
to fold the events in the ROI into 15 uniform orbital phase-bins (the zero point is set at MJD 54,682.66).
The phase-resolved spectral properties and fluxes within each phase-bin (Fig.~\ref{Fig2}) are obtained by running Tool \emph{gtlike} with the new model file.
We fit the phase-resolved light curve with a constant by employing the $\chi^2$-statistic and have $\chi^2/d.o.f=42/14$,
which indicates a significant variability. We have derived the phase-resolved SEDs
for phase-bins with $4/15<\phi<5/15$ and $13/15<\phi<14/15$ to examine the difference in high (phase-bin 5) and low states (phase-bin 14)
(see the blue and green vertical bars in the upper panel of Fig.~\ref{Fig2}, respectively).
The three SEDs (including the phase-averaged one) are shown in the A, B and C panels of {Appendix} Fig.~\ref{fig:phase-resolved SED}, respectively.
The lines represent he best-fit results derived by Tool \emph{gtlike}.
{In {Appendix} Table 1 we summarize our best-fit parameters {(the integrated energy flux and isotropic gamma-ray luminosity)} of phase-resolved SED (see {Appendix} Fig.~\ref{fig:phase-resolved SED}) of the gamma-ray emission in the ROI.
The Phase-bin$_{\rm All}$, Phase-bin$_{\rm H}$ and Phase-bin$_{\rm L}$  represent the time-averaged, high and low state phase-bins, respectively.}
The spectrum in low state is harder than in high state.
{The flux difference between ``high" and ``low" states is $\sim1.5\times10^{-8}$~photons~cm$^{-2}$~s$^{-1}$ (corresponding to a luminosity of $\sim2.80\times10^{34}~{\rm erg~s^{-1}}$).
And the spectrum in the ``low" state is harder than in the ``high" state (see {Appendix} Fig.~\ref{fig:phase-resolved SED} and {Appendix} Tab.~1).
We conjecture that the gamma-rays in the ``low" state come from a population of isolated MSPs, while the gamma-rays in the high state also consist of the peak (periodic) emission of the new source. As shown in for example \citet{XingY2015,XingY2016}, the emission from isolated MSPs are usually harder than that from Black Widows/Redback system. This may help explain
the softer spectrum of our signal in the high state.}

%{\bf The modulation period of our signal and some known MSP binaries.} We show in {Extended} Data Fig.~\ref{fig:distribution} the distribution of the radio/optical orbital periods for the {15 known MSP binaries},
%available at http://www.naic.edu/$\sim$pfreire/GCpsr.html, in 47 Tuc. Though our signal has a periodic period well within the region,
%but does not exactly match any known {MSP binaries} in 47 Tuc.

%**********************************Figure 1-2***************************************
\renewcommand{\figurename}{\bf Appendix Figure}
\setcounter{figure}{0}
\begin{figure}[!ht]
\centering
\includegraphics[scale=0.5]{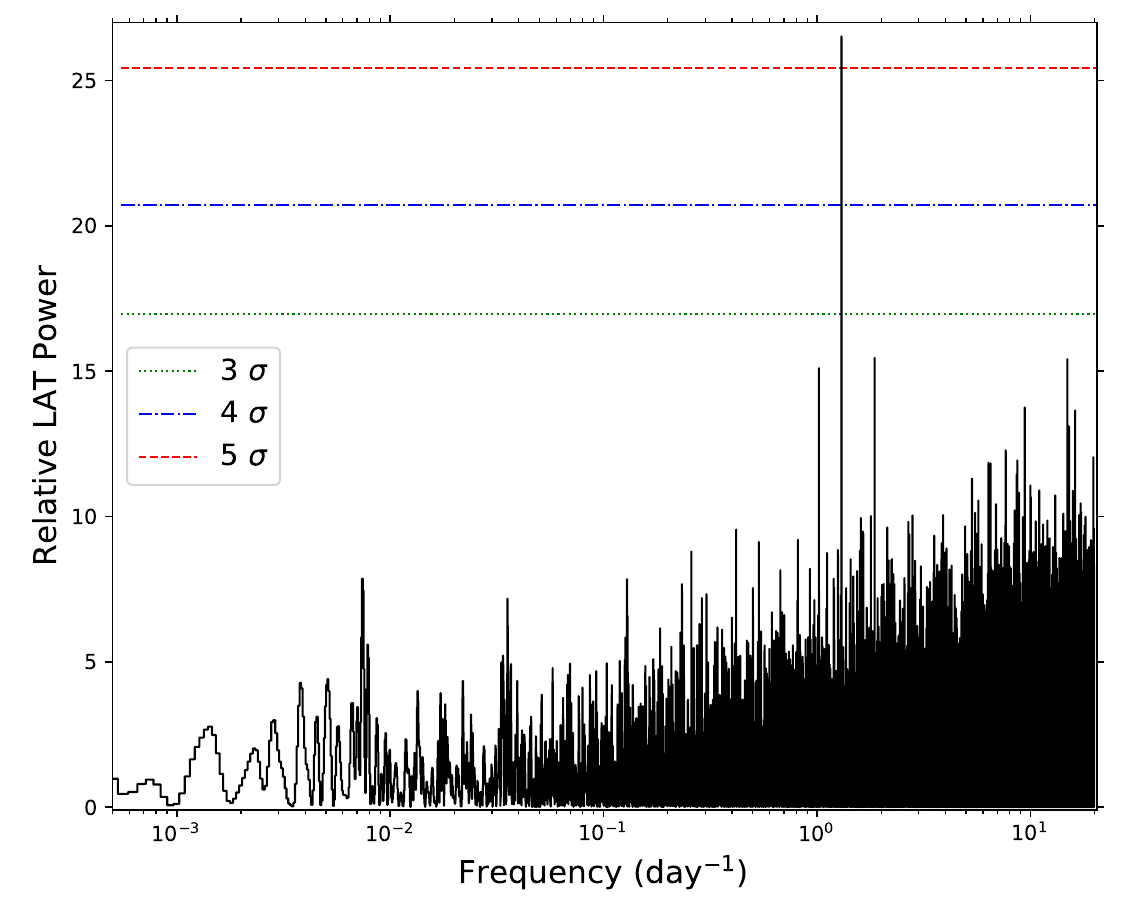}
\caption{Same as Fig.~\ref{Fig1}, but for power spectrum calculated from gamma-ray light curve with data up to $\sim$5 GeV.}
\label{Fig1-2}
\end{figure}
%%%%%%%%%%%%%%%%%%%%%%
%**********************************Figure 4***************************************
\renewcommand{\figurename}{\bf Appendix Figure}
\setcounter{figure}{1}
\begin{figure}[!ht]
\centering
\includegraphics[scale=0.33]{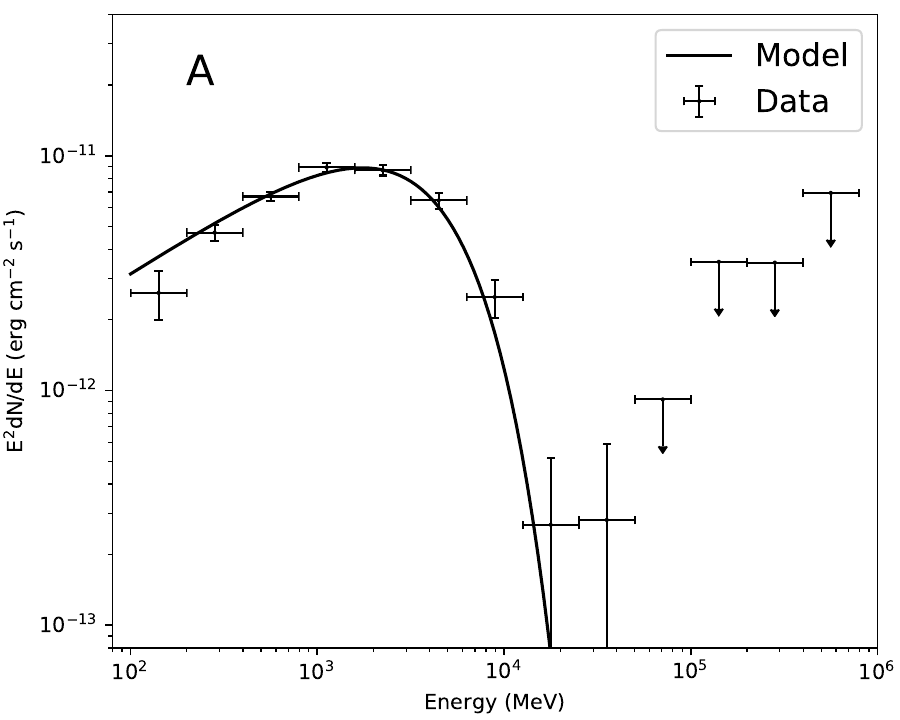}
\includegraphics[scale=0.33]{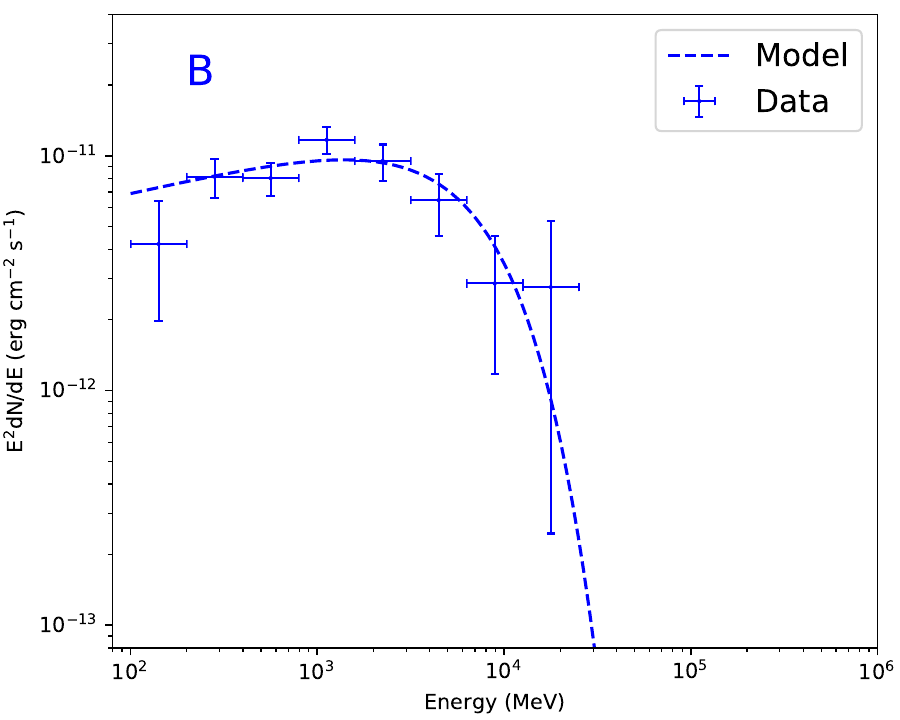}
\includegraphics[scale=0.33]{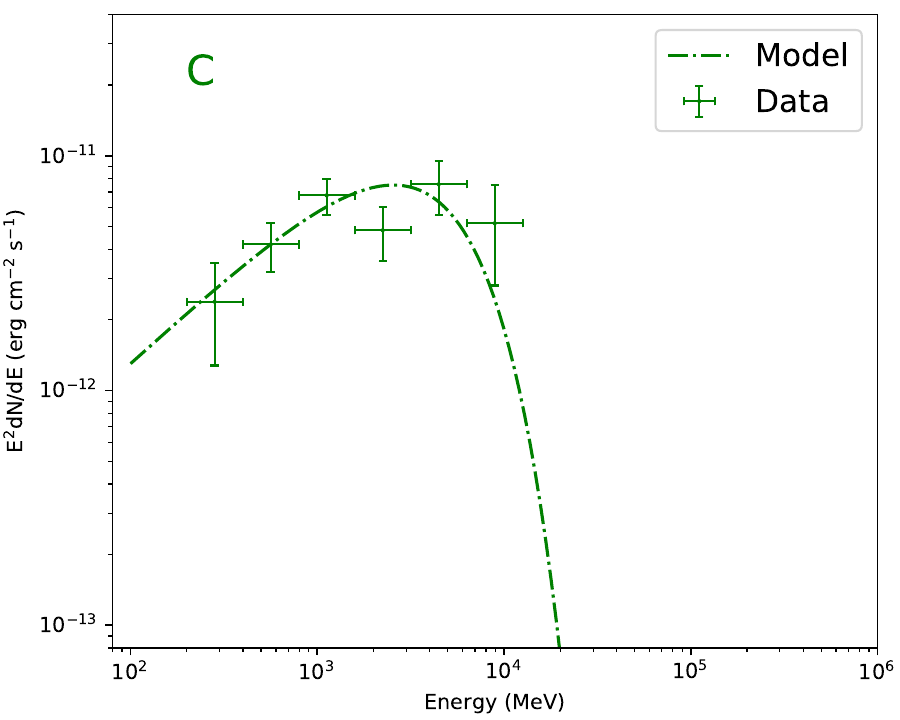}
\caption{Phase-averaged SED and the ones in high and low state phase-bins are shown in A, B, and C panel, respectively,
              with the lines indicating the best-fit spectral models in likelihood analysis presented in Tab.~1.
              {Note that, for panel A, the TS value of the data point with large error in 25-50 GeV is $\sim$4.}}
\label{fig:phase-resolved SED}
\end{figure}
%%%%%%%%%%%%%%%%%best-fit%%%%%%%%%%%%%%
\begin{table}
\begin{center}
\title{}{\bf Appendix Table 1}. {Results of phase-resolved analysis.}\\
\begin{tabular}{{c|c|c|c|c|c|c}}
\hline\hline
Phase-bin                     & $\Gamma$      &  b  &  $E_{\rm cut}$ (GeV)  &  TS value & $f_\gamma$ ($\times10^{-11}$ ergs/cm$^2$/s) & $L_\gamma$ ($\times10^{34}$ ergs/s)\\ [0.01cm]
\hline
Phase-bin$_{\rm All}$        &  $1.48\pm0.12$        &  $1.14\pm0.24$  &  $3.5\pm1.1$  &  6196.5 & 3.01 $\pm$ 0.08 & 5.76 $\pm$ 0.16  \\
Phase-bin$_{\rm H}$         &  $1.82\pm0.13$        &  1.14$^{\rm(\dag)}$  &  $6.9\pm2.8$  &  538.7 & 3.55 $\pm$ 0.27 &6.79 $\pm$ 0.52   \\
Phase-bin$_{\rm L}$         &  $1.26\pm0.25$        &   1.14$^{\rm(\dag)}$ &  $3.7\pm1.3$  &  296.3  & 2.09 $\pm$ 0.21 &3.99 $\pm$ 0.41  \\
\hline\hline
\end{tabular}
\end{center}
{{\bf Note:} {Best-fit parameters of phase-resolved likelihood analysis for the time-averaged, high and low state phase-bins.
              $b$ marked with $^{\dag}$ are fixed in the likelihood analysis for high and low states phase-bins. {$f_\gamma$ is integrated energy flux.
              $L_\gamma$ is isotropic gamma-ray luminosity.}}}
\label{best-fit}
\end{table}
%Our gamma-ray modulation period $\sim18.4$ hours is indeed within such a range (see {Appendix} Fig.~\ref{fig:distribution}).
%%**********************************Figure 5***************************************
%\renewcommand{\figurename}{\bf Appendix Figure}
%\setcounter{figure}{2}
%\begin{figure}[!ht]
%\centering
%%\includegraphics[scale=0.8]{E_Fig3.pdf}
%\caption{The distribution of the radio/optical orbital periods for the 15 pulsar binary systems in 47 Tuc.
%              Green dashed line indicates the gamma-ray periodic period reported in this work.}
%\label{fig:distribution}
%\end{figure}
%%%%%%%%%%%%%%%%%%
\renewcommand{\figurename}{\bf Appendix Figure}
\setcounter{figure}{2}
\begin{figure}[!ht]
\centering
\includegraphics[scale=0.4]{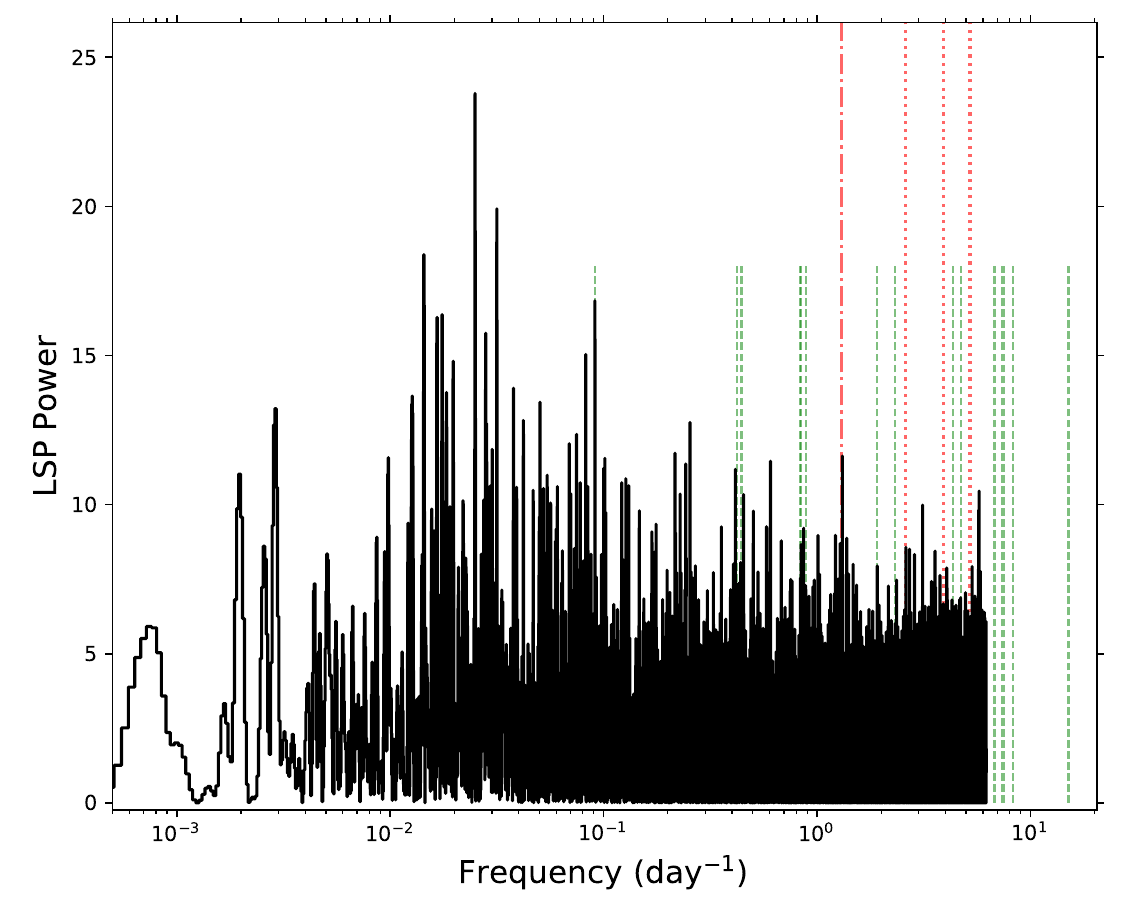}
\caption{Power spectrum of the \emph{Swift}-BAT X-ray light curve of 47 Tuc. The red dashed-dotted line indicates the period detected in gamma-ray{, and the red dotted lines are its second, third, and fourth harmonics, respectively}.
              The shorter green dashed lines indicate the periods of currently 15 known MSP binaries {(acquired at http://www.naic.edu/$\sim$pfreire/GCpsr.html)} in 47 Tuc.}
\label{Fig3}
\end{figure}
%%%%%%
\renewcommand{\figurename}{\bf Appendix Figure}
\setcounter{figure}{3}
\begin{figure}[!ht]
\centering
\includegraphics[scale=0.25]{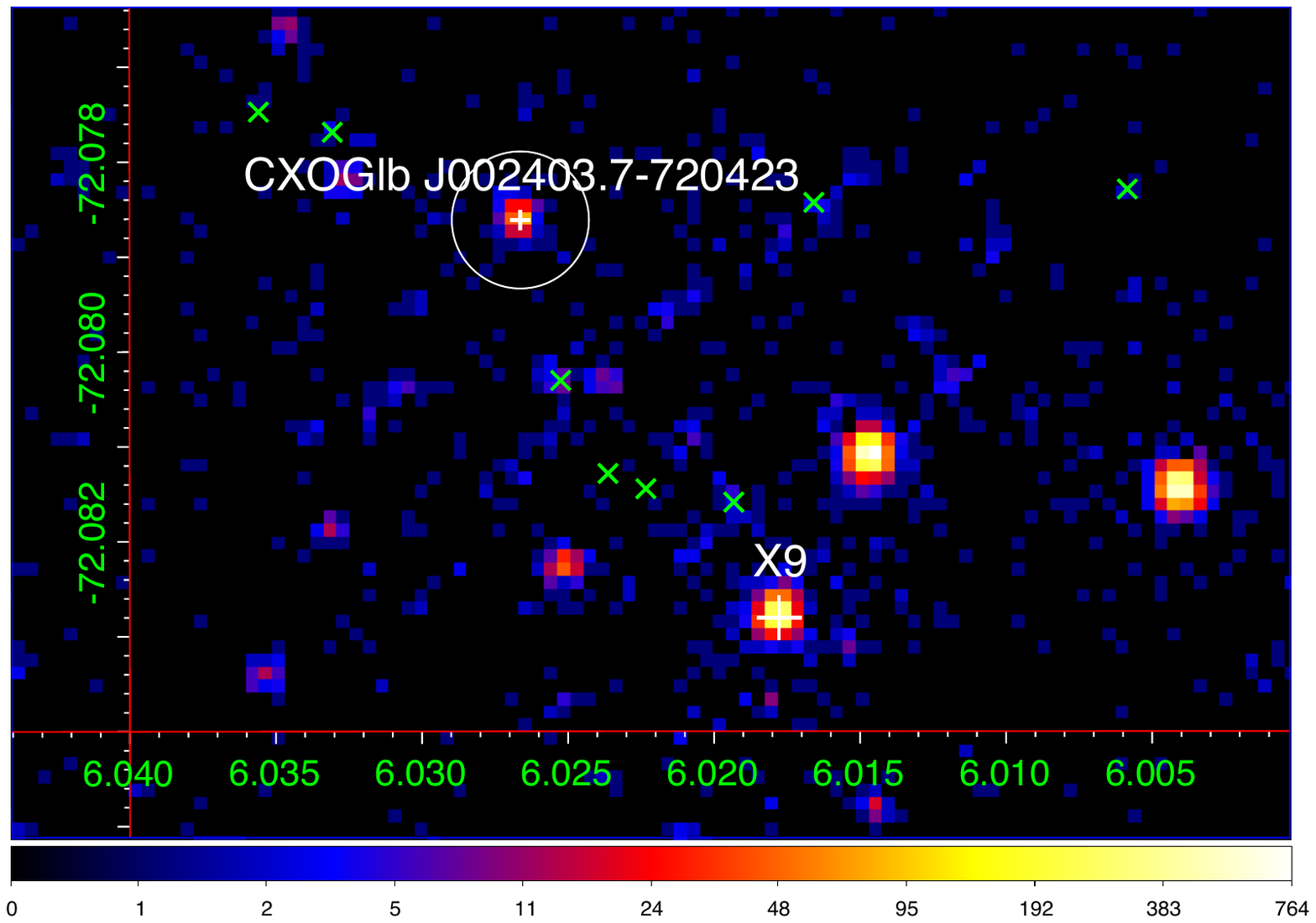}
\includegraphics[scale=0.25]{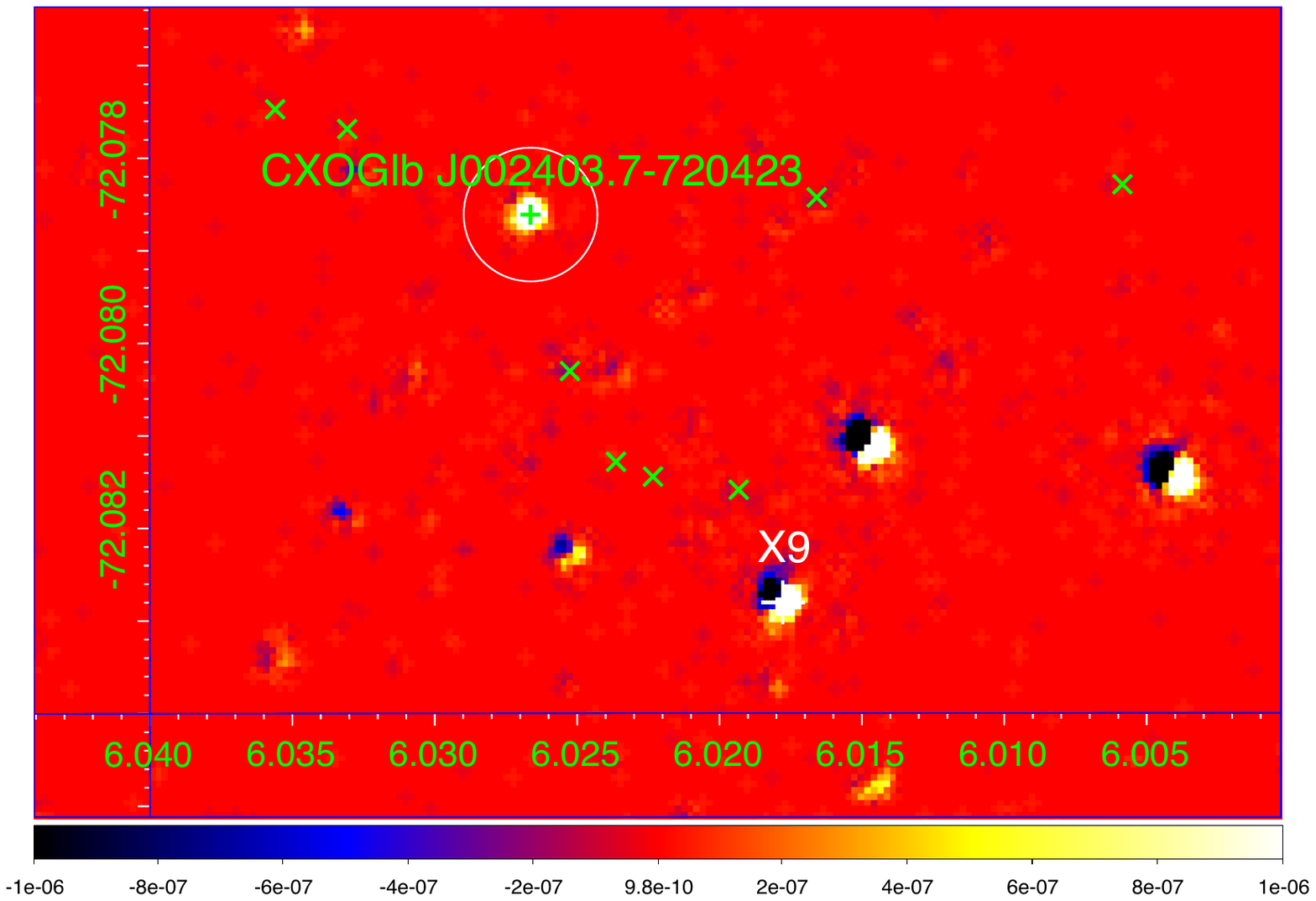}
\caption{{Counts map created with the data of ObsID 955 (left) and the residual map created with the data of ObsID 955 and 953 (right). The candidate source (CXOGlb J002406.3-720443) is marked with a cross.}
              A black hole candidate X-ray binary 47 Tuc X9 (R. A. = $00^{\rm h}24^{\rm m}04^{\rm s}.264$ and decl. = $-72^{\circ}04'58''.09$) \citep{Bahramian2017,Miller-Jones2015}
              is also marked. And the green ``$\times$" indicate the {MSPs} that have been identified in 47 Tuc.}
\label{Fig4}
\end{figure}
%%%%%%%
%\renewcommand{\figurename}{\bf Appendix Figure}
%\setcounter{figure}{5}
%\begin{figure}[!ht]
%\centering
%%\includegraphics[scale=0.5]{E_Fig6.pdf}
%\caption{The power spectrum of the merged light curve of the candidate source. The data were collected from 2002 September 29 and 2002 October 11, in total four observations (with the ID number of 2735, 2736, 2737, and 2738). Each observation has an exposure time about 65 ks. {And the green dashed line represents the 99\% confidence level}.}
%\label{Fig5a}
%\end{figure}
%%%%%%%%%%%%%%%%%%%%%%%

{\bf X-ray emission analysis.} To verify our gamma-ray periodic signal and identify the candidate source, we have collected the  X-ray archive data of 47 Tuc.
{A long-term set is provided by the {\it Swift}/Burst Alert Telescope (BAT) hard X-ray transient monitor \citep{Barthelmy2005,Krimm2013} spaning from MJD 53,416 to 58,134.
%Such a data set is available at https://swift.gsfc.nasa.gov/results/transients/weak/NGC104.orbit.lc.txt.
%\footnote{https://swift.gsfc.nasa.gov/results/transients/weak/NGC104.orbit.lc.txt}
Its power spectrum is presented in {Appendix} Fig.~\ref{Fig3}.
Interestingly, a peak with a local significance of $\sim 4.5\sigma$ (the probability is 99.9992\%) is found almost at the same period as that found in gamma-rays (see the red dashed-dotted line).}
%This peak has a local significance level of $\sim 4.5\sigma$ (99.9992\%).
Such a consistence provides strong support to the periodic modulation signal shown in the \emph{Fermi}-LAT of 47 Tuc.
Since the {\it Swift}/BAT data cannot provide accurate location information of the candidate source(s), below we focus on
the Chandra \citep{Fruscione2006} X-ray data that are characterized by the unprecedented angular resolution \citep{Weisskopf2002}.

%The archive Chandra X-ray data are available at http://cda.harvard.edu/chaser.
To identify the possible candidates for the gamma-ray modulation with the period of $\sim 18$ hours,
we create two exposure-corrected images of the observations with the ID number of 955 and 953 respectively.
Their time interval is both $\sim$ 10 hours (i.e., about half of the modulation period of our signal).
{If our gamma-ray periodic modulation is caused by the orbital period of a binary, these two exposure-corrected images (having same exposure time $\sim$8.8 hour) should cover the flux maximum and minimum of the target, respectively.}
The digital subtraction is carried out to get the residual map {(shown in the right panel of {Appendix} Fig.~\ref{Fig4})}, with which we find five bright sources.
%Note that the Chandra data of 47 Tuc are not dense enough and the longest individual observation {(from ACIS-I and ACIS-S detectors)} just lasted less than 20 hours. Therefore,
%the power spectrum analysis result is less reliable. We instead get the phase-folded light curves and check the consistence with the gamma-ray signal.
Our best candidate is spatially consistent with CXOGlb J002406.3-720443 {shown in {Appendix} Fig.~\ref{Fig4} marked with a cross.
%Such a candidate is marked with a white cross in {Appendix} Fig.~\ref{Fig4}.
Its X-ray light curves (0.3-8 keV) are extracted from a circle ROI of $\sim2.6$ arcsec radius with time bin of 4.4 ks for all observations from Chandra ACIS detectors.}
We fold them with a period cycle of 18.4 hours (zero phase is set at MJD 54,682.66) and present the resulting light curve in the lower panel of Fig.~\ref{Fig2}.
Its averaged X-ray energy flux is $(1.29\pm0.22)\times10^{-13}$ ergs cm$^{-2}$ s$^{-1}$. The gamma-ray-to-X-ray flux ratios are 233$\pm$40, 275$\pm$52, and 162$\pm$33 for the time-averaged, high and low states, respectively.
%For PSR J1311-3430 \citep{Arumugasamy2015,Pletsch2012}, the averaged X-ray and gamma-ray energy fluxes are $\sim8.6\times10^{-14}$ and $\sim6.2\times10^{-11}$ ergs cm$^{-2}$ s$^{-1}$;
%and its gamma-ray-to-X-ray flux ratio is $\sim720$. For PSR B1957-20 \citep{Huang2007,Abdo2010b,Guillemot2012},
%the averaged X-ray and gamma-ray energy fluxes are $\sim8.35\times10^{-14}$
%and $\sim1.34\times10^{-11}$ ergs cm$^{-2}$ s$^{-1}$; and hence its gamma-ray-to-X-ray flux ratio is $\sim160$.
%Therefore the gamma-ray-to-X-ray flux ratios reported here are within the range of ratios of PSR J1311-3430 and PSR B1957-20.}
Finally we would like to remark that the nature of CXOGlb J002406.3-720443 is still unclear. It was suggested to be a cataclysmic variable,
but a LMXB can not be ruled out, yet \citep{Grindlay2001}. %\del{???Either of them strongly suggests that there is a new kind of gamma-ray phenomena in Global Cluster.???}

%%%%%%%%%%%%%%%%%%%%%%%%
%%%%%%%%%%%%%%%%%%%%%%%%

%%%%%%%%%%%%%%%%%%%%%%%%%%%
\end{document}